\documentclass[twocolumn,10pt,jmp]{revtex4-1}
\usepackage{bm}
\usepackage{amsfonts}
\usepackage{amssymb}
\usepackage{amsmath}
\usepackage{graphicx}
\begin{document}
\title{Reduction of the classical electromagnetism to a two-dimensional curved surface}
\author{Tomasz Radozycki}
\email{torado@fuw.edu.pl}
\affiliation{Center for Theoretical Physics, Polish Academy of Sciences, Al. Lotnik\'ow 32/46, 02-668 Warsaw, Poland}

\begin{abstract}
The reduction of the three-dimensional classical electromagnetism is performed in a twofold way. In the first case the ordinary two-dimensional electromagnetism is obtained with sources in the form of conserved electric currents flowing along the surface. The electric field is a two-vector tangent to the surface and magnetic field is a scalar quantity. In the second approach the reduced theory is that of the two-vector magnetic field and a scalar electric one. The only source coupled to the fields is now a scalar subject to no conservation law. In the redefined theory this scalar source is may be converted into an eddy magnetic current flowing in the surface. No magnetic monopoles appear. Our results can find some applications in the electrodynamics of thin layers and of metal-dielectric interfaces. 
\end{abstract}
\maketitle

\section{Introduction}\label{intro}

Electromagnetic waves trapped in thin surfaces have long attracted the attention of both physicists and engineers due to various possible applications in optoelectronics, optical communication or integrated optics \cite{tam1,hun,tam2}.
One of possible practical realizations are thin dielectric layers of various shapes confining waves due to the phenomenon of total internal reflection and leading to the so called open waveguides~\cite{marcuse,sc,yeh,bal, carson,chew,collin}. While appropriately adjusting parameters as layer thickness, refractive index of the material, wave frequency, one can create a system in which only fundamental $TE$ and $TM$ modes are captured, all other being of radiative character and escaping from the layer. Depending on the electromagnetic properties of the media either $TE$ (dielectric layer) or $TM$ mode (magnetic layer) may be squeezed around the layer~\cite{brb}. 

Still more `flat' waves appear as the so called surface plasmon polaritons (SPPs), propagating along the metal-dielectric interface (for instance silver-silica or even silver-air) and being spatially confined in the perpendicular direction~\cite{rae}. Contrary to a typical dielectric waveguide, where fields are of evanescent character only outside of the border surface, SPPs exponentially decay on both sides of an interface: on one side because of the conductor properties and total internal reflection on the other. Their emergence results from the negative value of the dielectric constant (see for instance Drude-Sommerfeld model of electron gas~\cite{drude,som,as}) typical for the interaction of noble metal nanostructures with electromagnetic waves at optical frequencies~\cite{nh}. Practically such a system is created by placing a thin metallic film on a dielectric substrate. 

The description of light propagation in thin structures becomes more and more important due to increasing miniaturization in optoelectronics and intensive development of nanophysics. For this description two-dimensional reduced theory of electromagnetism~\cite{hillion,lapidus,zwiebach} is often used as a model. The situation in an open wave-guide to a certain extent may be mimicked in that way but with certain limitations~\cite{brb}. On the other hand in the case of SPPs the $TM$ modes are the only ones that propagate along the surface (corresponding $TE$ modes do not arise due to the boundary conditions), and cannot be described by the usual form of the $2D$ electromagnetism with the Lagrangian
\begin{equation}
L=-\frac{1}{4}\, F_{\mu\nu}F^{\mu\nu}+J_\mu A^\mu,
\label{l2d}
\end{equation}
where
\begin{equation}
F^{\mu\nu}=\left[\begin{array}{ccc} 0 & E_x & E_y\\ -E_x & 0 & B\\ -E_y & -B & 0\end{array}\right].
\label{fmn}
\end{equation}
Here and below we use the system of units, where $c=\hbar=1$. In this system the vacuum permeability is the inverse of permittivity:
$$
\mu_0=(\epsilon_0)^{-1}.
$$

In the present paper we would like to perform the reduction of the Maxwell electromagnetism from $3$ spatial dimensions to $2$ in a systematic way. In section~\ref{reduction} we show that this reduction can be done in a twofold way, leading to two versions of the Lagrangian: either that with electromagnetic tensor given by~(\ref{fmn}) or the one with electric and magnetic fields interchanged. This second version is more convenient to describe SPPs, and both are necessary to deal with the propagation in thin dielectric/magnetic layers. It is also needed in order to preserve $E\leftrightarrow B$ duality known from $3D$ theory (in the absence of sources).

The most convenient approach to perform this reduction is that coordinate independent, based on differential forms formulation of electromagnetism~\cite{desch,warnick}, since it allows to treat flat and curved surfaces on an equal footing. This language has the advantage (over tensor notation in its full complexity) of significantly simplifying calculations.

In section~\ref{sources}, we analyze the role of the scalar source appeared during reduction. We show, that this kind of a source can be connected with surface plasmons.

\section{Two-dimensional reduction of electromagnetism}\label{reduction}

In four space-time dimensions the set of Maxwell equations reads:
\begin{subequations}\label{max}
\begin{align}
&{\bm \nabla}\times {\mathbf H}=\partial_t{\mathbf D}+{\bm j},\label{max1}\\
&{\bm \nabla}\times {\mathbf E}=-\partial_t{\mathbf B},\label{max2}\\
&{\bm \nabla}\cdot{\mathbf D}=\rho,\label{max3}\\
&{\bm \nabla}\cdot{\mathbf B}=0.\label{max4}
\end{align}
\end{subequations}

The electromagnetic fields can be consistently coupled only to a conserved current, i.e. satisfying the continuity equation
\begin{equation}
\partial_t\rho+{\bm \nabla}\cdot{\bm j}=0,
\label{cont}
\end{equation}
which is the obvious consequence of equations~(\ref{max1}) and~(\ref{max3}).

For our applications it is especially convenient to rewrite these equations in terms of differential forms as:
\begin{subequations}\label{maxf}
\begin{align}
&{\mathrm d}\stackrel{1}{H}=\partial_t\stackrel{2}{D}+\stackrel{2}{J},\label{maxf1}\\
&{\mathrm d}\stackrel{1}{E}=-\partial_t \stackrel{2}{B},\label{maxf2}\\
&{\mathrm d}\stackrel{2}{D}=\stackrel{3}{\rho},\label{maxf3}\\
&{\mathrm d}\stackrel{2}{B}=0,\label{maxf4}
\end{align}
\end{subequations}
and
\begin{equation}
\partial_t\stackrel{3}{\rho}+{\mathrm d}\stackrel{2}{J}=0,
\label{contf}
\end{equation}
which are coordinate-independent and suitable for the reduction of electromagnetism to lower-dimensional curved surfaces. We adopt here the mathematical notation, in which the degree of a given form is marked with an index above its symbol. When this mark is absent, the symbol refers to the corresponding physical quantity and not the differential form.

Apart from (\ref{maxf1})-(\ref{maxf4}) there are also two constitutive relations:
\begin{subequations}\label{DEBH}
\begin{align}
&\stackrel{2}{D}=\epsilon_0*\stackrel{1}{E},\label{DE}\\
&\stackrel{2}{B}=\mu_0*\stackrel{1}{H},\label{BH}
\end{align}
\end{subequations}
specifying the connection between field intensities $\mathbf E$, $\mathbf H$ and flux densities  $\mathbf D$, $\mathbf B$ or rather their corresponding differential forms.

The Hodge star operator in a certain $N$-dimensional space (in (\ref{DE}) and (\ref{BH}) we mean 3D space, not the 4D space-time) acts on the base $k$-forms in the standard way:
\begin{eqnarray}
&*{\mathrm d} x_1\wedge\ldots\wedge {\mathrm d} x_k=\label{Hstar}\\
&\frac{1}{(n-k)!}\, \sqrt{g}\, g^{i_1j_1}\cdot\ldots\cdot g^{i_kj_k}\epsilon_{j_1,\ldots ,j_N}{\mathrm d} x_{k+1}\wedge\ldots\wedge {\mathrm d} x_N,
\nonumber
\end{eqnarray}
where $\epsilon_{j_1,\ldots ,j_N}$ denotes the $N$-dimensional antisymmetric symbol and $g$ is the determinant of the metric tensor $\hat{g}$. We use the usual convention, where matrix elements of $\hat{g}$ are denoted by lower indexes, i.e. $g_{ij}$, and the symbol $g^{ij}$ refers to the elements of the inverse tensor:
\begin{equation}
g_{ij}g^{jk}=\delta_i^k,
\label{ginv}
\end{equation}
$\delta_i^k$ being the Kronecker delta. Below, in order to distinguish between the metric tensor in three and two dimensions we will write either $\hat{g}_{3D}$ or  $\hat{g}_{2D}$. The Hodge star in various spaces will be denoted as $*_3$ or $*_{3+1}$ for three-dimensional space or $3+1$ dimensional space-time respectively. Similar notation $*_2$ or $*_{2+1}$ refers to reduced space (or space-time).

Now we would like to reduce the whole set of equations to a certain two-dimensional, and in general curved surface. This reduction will be performed in the following way. We first choose in the $3D$ space a system of orthogonal coordinates well-fitting to the considered limiting surface. They will be called $u,v,w$, referred to also by indexes as $1, 2, 3$. The $w$ coordinate plays a special role: the corresponding tangent vector (to say it precisely: the tangent vector to the $3D$ space, which is, however, normal to the considered sub-surface) defines the direction along which the projection is going to be done. The additional condition we impose while choosing the coordinates is that $|\nabla_w {\mathbf r}|=1$ (we use the notation $\nabla_w=\partial/\partial w$). This means that this tangent vector may rotate, while changing $u$ and $v$, but it cannot change its length. Such a choice gives:
\begin{equation}
g_{33}=\nabla_w {\mathbf r}\cdot \nabla_w {\mathbf r}=1.
\label{gww}
\end{equation}

The connection with the Cartesian system is then given by the dependence ${\mathbf r}(u,v,w)$. 
In terms of these curvilinear coordinates the field forms read:
\begin{subequations}\label{ff}
\begin{align}
&\stackrel{1}{E}=E_u{\mathrm d} u+E_v{\mathrm d} v+E_w{\mathrm d} w\label{ef}\\
&\stackrel{2}{D}=D_u{\mathrm d} v\wedge {\mathrm d} w+D_v{\mathrm d} w\wedge {\mathrm d} u+D_w{\mathrm d} u\wedge {\mathrm d} v,\label{df}\\
&\stackrel{1}{H}=H_u{\mathrm d} u+H_v{\mathrm d} v+H_w{\mathrm d} w\label{hf}\\
&\stackrel{2}{B}=B_u{\mathrm d} v\wedge {\mathrm d} w+B_v{\mathrm d} w\wedge {\mathrm d} u+B_w{\mathrm d} u\wedge {\mathrm d} v,\label{bf}
\end{align}
\end{subequations}
and those for the source are
\begin{subequations}\label{current}
\begin{align}
&\stackrel{3}{\rho}=\rho\, {\mathrm d} u\wedge {\mathrm d} v\wedge {\mathrm d}  w,\label{current1}\\
&\stackrel{2}{J}=j_u{\mathrm d} v\wedge {\mathrm d} w+j_v{\mathrm d} w\wedge {\mathrm d} u+j_w{\mathrm d} u\wedge {\mathrm d} v.\label{current2}
\end{align}
\end{subequations}

This identification is consistent, since electric field is in fact a force acting on a unit charge and therefore in the natural way is associated with certain one-form (i.e the work form). Similarly $\mathbf H$ appears in magnetic circuit and should be linked to one-form too. In turn such quantities as $\mathbf B$ or $\mathbf D$ are connected with fluxes, so they correspond to two-forms.

In the literature the source forms are often defined as $\stackrel{0}{\rho}$ and $\stackrel{1}{J}$ and those given by (\ref{current}) are treated as $*_3\!\stackrel{0}{\rho}$ and $*_3\!\stackrel{1}{J}$ respectively. The definition (\ref{current}) seems, however, more natural, since $\rho$ is a volume and $J$ a surface density.

If the coordinates are correctly chosen, the surface is obtained by putting $w=0$ and is then parametrized by ${\mathbf r}(v,u,0)$. A unit vector normal to the surface at a given point may be written as
\begin{equation}
{\mathbf e}_n=\left. \frac{\nabla_w{\mathbf r}(u,v,w)}{|\nabla_w{\mathbf r}(u,v,w)|}\right|_{w=0}=\left. \nabla_w{\mathbf r}(u,v,w)\right|_{w=0}.
\label{normal}
\end{equation}

Now, the reduction to the two-dimensional space, which is parametrized by $u$ and $v$, may be accomplished in a twofold way.
First we take the {\em right} interior product (i.e. such that the contracted vector stands on the {\em right} of the form) of all equations~(\ref{maxf1}--\ref{maxf4}) with the normal vector ${\mathbf e}_n$, obtaining the following set:
\begin{subequations}\label{maxfo}
\begin{align}
&({\mathrm d}\stackrel{1}{H})\rfloor {\bm e}_n=\partial_t\stackrel{2}{D}\rfloor {\bm e}_n+\stackrel{2}{J}\rfloor {\bm e}_n,\label{maxfo1}\\
&(*_3{\mathrm d}\stackrel{1}{E})\rfloor {\bm e}_n=-(*_3\partial_t \stackrel{2}{B})\rfloor {\bm e}_n,\label{maxfo2}\\
&({\mathrm d}\stackrel{2}{D})\rfloor {\bm e}_n=\stackrel{3}{\rho}\rfloor {\bm e}_n,\label{maxfo3}\\
&(*_3{\mathrm d}\stackrel{2}{B})\rfloor {\bm e}_n=0,\label{maxfo4}
\end{align}
\end{subequations}

The presence of Hodge star operator in two equations will be justified later (see the comment after formula~(\ref{jsplit})).
Since in the chosen orthogonal coordinates ${\mathrm d} u\rfloor {\bm e}_n ={\mathrm d} v\rfloor {\bm e}_n=0$ and ${\mathrm d} w\rfloor {\bm e}_n=1$, we obtain:
\begin{subequations}\label{int}
\begin{align}
&{\mathrm d} u\wedge {\mathrm d} v\rfloor {\bm e}_n=0,\;\;\; {\mathrm d} w\wedge {\mathrm d} u\rfloor {\bm e}_n=-{\mathrm d} u,\nonumber\\
\label{int1}\\ 
&{\mathrm d} v\wedge {\mathrm d} w\rfloor {\bm e}_n={\mathrm d} v,\;\;\; {\mathrm d} u\wedge {\mathrm d} v\wedge {\mathrm d} w\rfloor {\bm e}_n={\mathrm d} u\wedge {\mathrm d} v,\nonumber\\
\label{int2}
\end{align}
\end{subequations}

The above set leads (upon neglecting the $w$ derivative since no quantity in our `flat' world may depend on this variable) to the following Maxwell equations in two dimensions:
\begin{subequations}\label{me}
\begin{align}
&\nabla_v H=\partial_t D_u+{j_{e}}_u,\label{me1}\\
&\nabla_u H=-\partial_t D_v-{j_{e}}_v,\label{me2}\\
&\nabla_uE_v-\nabla_vE_u=-\partial_t B,\label{me3}\\
&\nabla_uD_u+\nabla_vD_v=\rho_e,\label{me4}
\end{align}
\end{subequations}
all symbols being explained below. The additional subscript $e$ stands for `electric' to distinguish the sources from magnetic ones dealt with below.  The Eq.~(\ref{maxfo4}) is omitted, since it is an identity. 

The current satisfies the two-dimensional version of the continuity equation:
\begin{equation}
\partial_t\rho_e+\nabla_u{j_{e}}_u+\nabla_v {j_{e}}_v=0,
\label{cont2}
\end{equation}
which results from~(\ref{me1}), (\ref{me2}) and~(\ref{me4}) and could also have been written as
\begin{equation}
\partial_t \stackrel{3}{\rho}\rfloor {\bm e}_n+({\mathrm d}\stackrel{2}{J})\rfloor {\bm e}_n=0.
\label{conr}
\end{equation}

The reduced constitutive equations in the considered case have the form
\begin{subequations}\label{DrBr}
\begin{align}
&\stackrel{2}{D}\rfloor {\bm e}_n=\epsilon_0(*_3\!\stackrel{1}{E})\rfloor {\bm e}_n,\label{DEr}\\
&(*_3\!\stackrel{2}{B})\rfloor {\bm e}_n=\mu_0\stackrel{1}{H}\rfloor {\bm e}_n,\label{BHr}
\end{align}
\end{subequations}
since in three dimensions $**$ is an identity. The value of $\epsilon_0$ in $2D$ is in general different from that in $3D$ and should be determined from the measurement of the Coulomb force in such a hypothetic  world. The relations (\ref{DEr}) and (\ref{BHr}) expressed in terms of coordinates become:
\begin{subequations}\label{DvBv}
\begin{align}
&D_u=\epsilon_0\sqrt{g_{3D}}\,(g_{3D})^{11}E_u,\label{DEu}\\
&D_v=\epsilon_0\sqrt{g_{3D}}\,(g_{3D})^{22}E_v,\label{DEv}\\
&B=\mu_0\sqrt{g_{3D}\,}H,\label{BHw}
\end{align}
\end{subequations}
where the symbol $g_{3D}$ here and below naturally refers to $g_{3D}|_{w=0}$.

The above reduced equations should be compared to what one obtains while directly formulating the electromagnetism on the two-dimensional surface. Instead of (\ref{hf})-(\ref{bf}) and (\ref{current}) we would, then, write:
\begin{subequations}\label{f2d}
\begin{align}
&\stackrel{1}{E}=E_u{\mathrm d} u+E_v{\mathrm d} v\label{ef2di}\\
&\stackrel{1}{D}=D_u{\mathrm d} v-D_v {\mathrm d} u,\label{df2di}\\
&\stackrel{0}{H}=H\label{hf2di}\\
&\stackrel{2}{B}=B{\mathrm d} u\wedge {\mathrm d} v\label{bf2di}
\end{align}
\end{subequations}
and also
\begin{equation}
\stackrel{2}{\rho}=\rho_e\, {\mathrm d} u\wedge {\mathrm d} v,\;\;\; \stackrel{1}{J}={j_{e}}_u{\mathrm d} v-{j_{e}}_v{\mathrm d} u.
\label{currenta}
\end{equation}

The minus sign in ~(\ref{df2di}) is due to the application of the Hodge star operation resulting from constitutive equation. The Maxwell and continuity equations have the form analogous to (\ref{maxf1})-(\ref{maxf4}) and (\ref{contf}), with the obvious modifications concerning the degree of the corresponding differential forms. The Eq.~(\ref{maxf4}) becomes here an obvious identity. Written in terms of variables $u$ and $v$, these equations turn out to be identical to (\ref{me1})-(\ref{me4}) and (\ref{cont2}). Some attention, however, should paid to the constitutive equations. In $2D$, we obtain:
\begin{subequations}\label{DBvu}
\begin{align}
&D_u=\epsilon_0\sqrt{g_{2D}}\,(g_{2D})^{11}E_u,\label{DEu2}\\
&D_v=\epsilon_0\sqrt{g_{2D}}\,(g_{2D})^{22}E_v,\label{DEv2}\\
&B=\mu_0\sqrt{g_{2D}}H,\label{BHw2}
\end{align}
\end{subequations}

In orthogonal coordinates we have (there is no summation over $i$ in the formula below):
\begin{eqnarray}
\sqrt{g_{3D}}\,(g_{3D})^{ii}=&\!\!\!\sqrt{(g_{3D})_{33}}\sqrt{g_{2D}}\,(g_{2D})^{ii}\nonumber\\
=&\,\sqrt{g_{2D}}\,(g_{2D})^{ii},\;\;\; i=1,2,
\label{tm32}
\end{eqnarray}
where the last equality is owed to (\ref{gww}). Now we see, that the constitutive equations (\ref{DEu2})-(\ref{BHw2}) become identical to (\ref{DEu})-(\ref{BHw}), if we forget the different values and units of the permittivity constant $\epsilon_0$. 

The second alternative for the reduction, in place of~(\ref{maxfo1})-(\ref{maxfo4}), consists on considering 
\begin{subequations}\label{maxfoi}
\begin{align}
&(*_3{\mathrm d}\stackrel{1}{H})\rfloor {\bm e}_n=(*_3\partial_t\stackrel{2}{D})\rfloor {\bm e}_n+(*_3\!\stackrel{2}{J})\rfloor {\bm e}_n,\label{maxfoi1}\\
&({\mathrm d}\stackrel{1}{E})\rfloor {\bm e}_n=-\partial_t \stackrel{2}{B}\rfloor {\bm e}_n,\label{maxfoi2}\\
&(*_3{\mathrm d}\stackrel{2}{D})\rfloor {\bm e}_n=(*_3\!\stackrel{3}{\rho})\rfloor {\bm e}_n,\label{maxfoi3}\\
&({\mathrm d}\stackrel{2}{B})\rfloor {\bm e}_n=0,\label{maxfoi4}
\end{align}
\end{subequations}
equivalent to
\begin{subequations}\label{mm}
\begin{align}
&\nabla_uH_v-\nabla_vH_u=\partial_t D+j_m,\label{mm1}\\
&\nabla_u E=\partial_t B_v,\label{mm2}\\
&\nabla_v E=-\partial_t B_u,\label{mm3}\\
&\nabla_uB_u+\nabla_vB_v=0,\label{mm4}
\end{align}
\end{subequations}
where subscript $w$ of $j$ has been omitted and $m$ is added to mark the connection of the scalar $j$ with magnetic sources.

Note that the continuity equation
\begin{equation}
(*_3\partial_t \stackrel{3}{\rho})\rfloor {\bm e}_n+(*_3{\mathrm d}\stackrel{2}{J})\rfloor {\bm e}_n=0.
\label{conrs}
\end{equation}
does not introduce any new condition, since it becomes an identity (any $0$-form cannot be contracted with a vector). Therefore, no restrictions on $j$ arise. 

The reduced constitutive equations in this case are
\begin{subequations}\label{DBrq}
\begin{align}
&(*_3\!\stackrel{2}{D})\rfloor {\bm e}_n=\epsilon_0\stackrel{1}{E}\rfloor {\bm e}_n,\label{DErq}\\
&\stackrel{2}{B}\rfloor {\bm e}_n=\mu_0(*_3\!\stackrel{1}{H})\rfloor {\bm e}_n,\label{BHrq}
\end{align}
\end{subequations}
and in terms of coordinates they read:
\begin{subequations}\label{DBwvu}
\begin{align}
&\sqrt{g_{3D}}\,(g_{3D})^{11}(g_{3D})^{22}D=\epsilon_0E,\label{DEwq}\\
&B_u=\mu_0\sqrt{g_{3D}}\,(g_{3D})^{11}H_u,\label{BHuq}\\
&B_v=\mu_0\sqrt{g_{3D}}\,(g_{3D})^{22}H_v.\label{BHvq}
\end{align}
\end{subequations}

Since in the chosen variables we have $(g_{3D})^{33}=1$
and $ (g_{3D})^{ii}=[(g_{3D})_{ii}]^{-1}$, then 
\begin{equation}
\sqrt{g_{3D}}\,(g_{3D})^{11}(g_{3D})^{22}=\frac{1}{\sqrt{g_{3D}}},
\label{tms}
\end{equation}
and the relation (\ref{DEwq}) between $D$ and $E$ may be given the more common form:
\begin{equation}
D=\epsilon_0\sqrt{g_{3D}}\,E.\label{DEwqt}
\end{equation}

All obtained equations agree with the two-dimensional ones only if we formulate the electromagnetism via usual Maxwell equations with:
\begin{subequations}\label{dhb}
\begin{align}
&\stackrel{0}{E}=E\label{ef2d}\\
&\stackrel{2}{D}=D{\mathrm d} u\wedge{\mathrm d} v\label{df2d}\\
&\stackrel{1}{H}=H_u{\mathrm d} u+H_v{\mathrm d} v\label{hf2d}\\
&\stackrel{1}{B}=B_u{\mathrm d} v-B_v {\mathrm d} u\label{bf2d}
\end{align}
\end{subequations}
instead of~(\ref{f2d}).

The reduction of field equations to the $2$ dimensional space may be seen as equivalent to dividing components of the vector variables in the following way:
\begin{subequations}\label{desbhs}
\begin{align}
&{\mathbf E}\;\mapsto\; E_u, E_v, E; \;\;\; {\mathbf D}\;\mapsto\; D_u, D_v, D;\label{desplit}\\
&{\mathbf B}\;\mapsto\; B_u, B_v, B; \;\;\; {\mathbf H}\;\mapsto\; H_u, H_v, H,\label{bhsplit}
\end{align}
\end{subequations}
and similarly for the sources:
\begin{equation}
{\mathbf j}\;\mapsto\; {j_{e}}_u, {j_{e}}_v, j_m.
\label{jsplit}
\end{equation}
The subscript $w$ has been neglected everywhere, to emphasize that quantities $E,D,H,B,j_m$ are now scalars and not third components of vectors. At a given point there exist, then, two-dimensional vectors lying in the plane tangent to the surface and scalars. The set of Maxwell equations~(\ref{max1}--\ref{max4}) may be written in the form of two above disjoint subsets~(\ref{me1}--\ref{me4}) and (\ref{mm1}--\ref{mm4}). The fact that the whole set breaks up into two subsets (together with the constitutive equations) justifies the use of Hodge star in equations~(\ref{maxfo2}), (\ref{maxfo4}), (\ref{maxfoi1}), (\ref{maxfoi3}) spoken of below Eqs.~(\ref{maxfo}). 

While looking at (\ref{desplit}) and (\ref{bhsplit}) one should, however, remember that $\mathbf D$ and $\mathbf B$ have in $2D$ other dimensions than their $3D$ counterparts. In a flat world the flux densities are integrated over curves rather than over surfaces. The change of dimensionality of fields is accounted for by the similar change in $\epsilon_0$, which is also necessary to maintain the modified Coulomb law.

There are then two kinds of electromagnetisms in $2$ spatial dimensions as well as two kinds of sources (we neglect the possibility of the existence of magnetic sources containing monopoles): a current flowing in the surface, having vector character subject to the reduced continuity equation~(\ref{cont2}) and a scalar $j$ with no conservation restrictions. The former generates electric field tangent to the surface and a scalar magnetic field, and the latter a tangent magnetic field and an electric scalar.

In terms of electromagnetic potential, these results may be obtained in the following way. First we define the one-form in $4$ space-time dimensions:
\begin{equation}
\stackrel{1}{A}=-\Phi {\mathrm d} t +A_u{\mathrm d} u+A_v{\mathrm d} v+A_w{\mathrm d} w,
\label{potential}
\end{equation}
and next the two-form corresponding to the electromagnetic tensor $F_{\mu\nu}$:
\begin{equation}
\stackrel{2}{F}={\mathrm d} \stackrel{1}{A}=-{\mathrm d} t\wedge\stackrel{1}{E}  +\stackrel{2}{B}.
\label{F2}
\end{equation}
Now we apply either
\begin{equation}
(*_{3+1}\!\stackrel{2}{F})\rfloor {\bm e}_n= -(*_{3+1}\,{\mathrm d} t\wedge\stackrel{1}{E})\rfloor {\bm e}_n+(*_{3+1}\stackrel{2}{B})\rfloor {\bm e}_n,
\label{r1}
\end{equation}
which leads to
\begin{subequations}\label{r1v}
\begin{align}
&E_u=-\nabla_u\Phi-\partial_tA_u,\;\;\; E_v=-\nabla_v\Phi-\partial_tA_v,\label{r1va}\\
&B=\nabla_uA_v-\nabla_vA_u, \label{r1vb}
\end{align}
\end{subequations}
or
\begin{equation}
\stackrel{2}{F}\rfloor {\bm e}_n=-{\mathrm d} t\wedge \stackrel{1}{E}\rfloor {\bm e}_n+\stackrel{2}{B}\rfloor {\bm e}_n,
\label{r2}
\end{equation}
giving
\begin{equation}
E=-\partial_t A,\;\;\; B_u=\nabla_vA,\;\;\; B_v=-\nabla_uA.
\label{r2v}
\end{equation}
where the components of the potential have been divided similarly to~(\ref{desbhs}):
\begin{equation}
{\mathbf A}\;\mapsto\; A_u, A_v, A.
\end{equation}

The first theory is then described by the ordinary vector potential with components $[\Phi(t,v,u), A_v(t,v,u), A_u(t,v,u)]$, out of which only one is independent after gauge fixing and exploiting the Gauss law, and the second one by a single scalar field $A(t,v,u)$, with no gauge freedom and no Gauss law. In this latter case the Lagrangian of the free electromagnetism would have the form usual for the scalar field:
\begin{eqnarray}
L=&&\frac{\epsilon_0}{2}\int{\mathrm d} u{\mathrm d} v\sqrt{g_{2D}}\label{lagr2}\\
&&\times\left[(\partial_t A)^2-(g_{2D})_{11}(\nabla_u A)^2-(g_{2D})_{22}(\nabla_v A)^2 \right].
\nonumber
\end{eqnarray}

\section{Sources}\label{sources}

What requires some attention is the character of the scalar source in (\ref{mm1}). Waves described by the set (\ref{mm}) can be produced in thin layers for instance by charges performing oscillatory motion within the layer, perpendicular to it. As mentioned earlier this kind of a source in $2+1$-dimensional space-time does not have to satisfy any conservation law. The Eqs.~(\ref{mm}) do not require any continuity equation for $j_m$. It becomes clear if we imagine our surface as emerged in the larger $3D$ space. Among various (conserved) currents flowing in this space, a class of those perpendicular to the two-dimensional sub-surface, can be distinguished. Imagine a moving electric charge breaking through the surface. From the point of view of an inhabitant of this `flat' world, such a charge is first emerging and then dissolving leaving eventually a wave propagating along (or within) the surface.

It would be convenient to model such a source in the $2D$ surface, where perpendicular degrees of freedom are suppressed, without referring to any larger space. The possible practical realization are surface plasmons for which electromagnetic modes of this kind are the only propagating and confined ones~\cite{zayats}. In this case the quantity $j_m(u,v)$ may correspond to time-dependent local surface charge density oscillations.

Let us come back for a moment to the four-dimensional space-time and temporarily use Cartesian coordinates. The electric current can be decomposed onto the transverse (i.e. sourceless) and longitudinal (i.e. irrotational)  components in the following way:
\begin{equation}
{\mathbf j}={\mathbf j}_T+{\mathbf j}_L={\bm \nabla}\times {\mathbf V}-{\bm \nabla}\partial_t U,\;\;\;\; \rho=\Delta U,
\label{jlt}
\end{equation}
where time derivative of $U$ has been introduced for convenience. The continuity equation~(\ref{cont}) is then automatically satisfied. The spurious fourth component may be gauged away 
\begin{equation}
{\mathbf V}\longmapsto {\mathbf V}+{\bm \nabla}\phi
\label{gauge}
\end{equation}
since $\phi$ does not appear in any physical equations, and one can impose the gauge condition ${\bm \nabla}{\mathbf V}=0$. Substituting~(\ref{jlt}) into the Maxwell equations:
\begin{subequations}\label{smax}
\begin{align}
{\bm \nabla}\times {\mathbf H}&=\partial_t{\mathbf D}+{\bm \nabla}\times {\mathbf V}-{\bm \nabla}\partial_t U,\label{smax1}\\
{\bm \nabla}\times {\mathbf E}&=-\partial_t{\mathbf B},\label{smax2}\\
{\bm \nabla}\cdot{\mathbf D}&=\Delta U,\label{smax3}\\
{\bm \nabla}\cdot{\mathbf B}&=0.\label{smax4}
\end{align}
\end{subequations}
we see that both ${\mathbf V}$ and $U$ can be absorbed by the following redefinition of electromagnetic fields:
\begin{subequations}\label{red}
\begin{align}
&{\mathbf E}\longmapsto {\mathbf E}+\frac{1}{\epsilon_0}{\bm \nabla}U\label{rede}\\
&{\mathbf D}\longmapsto {\mathbf D}+{\bm \nabla}U,\label{redd}\\
&{\mathbf H}\longmapsto {\mathbf H}+{\mathbf V},\label{redh}\\
&{\mathbf B}\longmapsto {\mathbf B}+\mu_0{\mathbf V}.\label{redb}
\end{align}
\end{subequations}
In the language of differential forms we would write:
\begin{subequations}\label{redf}
\begin{align}
&\stackrel{1}{E}\;\longmapsto \stackrel{1}{E}+\frac{1}{\epsilon_0}{\mathrm d}\stackrel{0}{U}\label{redfe}\\
&\stackrel{2}{D}\;\longmapsto \stackrel{2}{D}+*_3{\mathrm d}\stackrel{0}{U},\label{redfd}\\
&\stackrel{1}{H}\;\longmapsto \stackrel{1}{H}+\stackrel{1}{V},\label{redfh}\\
&\stackrel{2}{B}\;\longmapsto \stackrel{2}{B}+\mu_0*_3\stackrel{1}{V},\label{redfb}\\
\end{align}
\end{subequations}
with obvious notation. After this change ${\mathbf V}$ and $U$ reappear as magnetic current, but still sourceless, i.e. without magnetic charges:
\begin{subequations}\label{tmax}
\begin{align}
{\bm \nabla}\times {\mathbf H}&=\partial_t{\mathbf D},\label{tmax1}\\
{\bm \nabla}\times {\mathbf E}&=-\partial_t{\mathbf B}-{\mathbf j}_m,\label{tmax2}\\
{\bm \nabla}\cdot{\mathbf D}&=0,\label{tmax3}\\
{\bm \nabla}\cdot{\mathbf B}&=\rho_m.\label{tmax4}
\end{align}
\end{subequations}
where
\begin{equation}
{\mathbf j}_m=\mu_0\partial_t{\mathbf V},\;\;\;\; \rho_m=-\mu_0{\bm \nabla}\cdot{\mathbf V}=0,
\label{jmd}
\end{equation}
due to the gauge condition.

\begin{figure*}
\begin{center}
\includegraphics[width= 0.8\textwidth,angle=0]{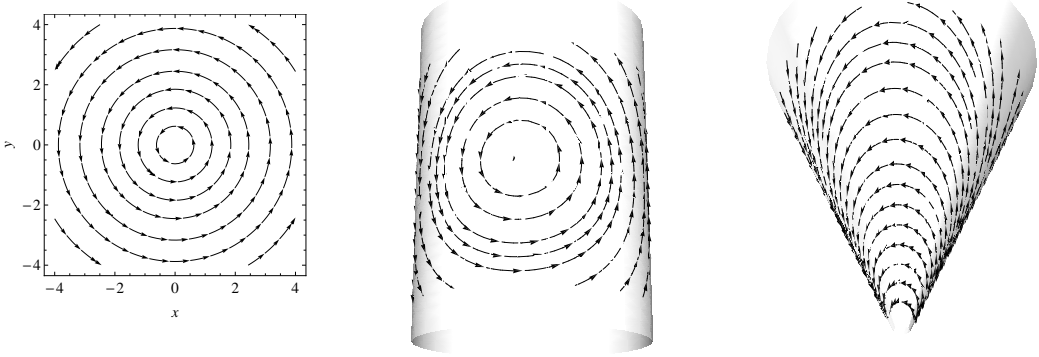}
\end{center}
\caption{Exemplary streams of the eddy current ${\mathbf j}_m$ on the plane, on a cylinder surface and on a cone surface in the gauge ${\bm \nabla}\cdot{\mathbf j}_m=0$. The arrows represent the field of vectors ${\mathbf j}_m(u,v)$, tangent to the surface.}
\label{streams}
\end{figure*}

Now we can turn to the scalar source appearing in~(\ref{mm1}). It would be natural to convert it into a current flowing within the two-dimensional surface. It can be done in the similar spirit as above. We will show, that this scalar source may be reinterpreted as a purely eddy magnetic current in the reduced space. 

The mathematical property of any smooth scalar function $j_m(u,v)$ is that it may be written in the following way:
\begin{equation}
j_m(u,v) =\nabla_uk_v(u,v) -\nabla_vk_u(u,v) 
\label{jk}
\end{equation}
where $k_u(u,v)$ and $k_v(u,v)$ are certain new functions, where again their time dependence has been omitted. The existence of them may be immediately justified if one associates a differential $2$-form with the scalar $j(u,v)$:
\begin{equation}
\stackrel{2}{j}=j_m(u,v){\mathrm d}u\wedge{\mathrm d}v.
\label{jf}
\end{equation}
As a $2$-form in $2$ dimensions this form is obviously closed and by virtue of Poincar\'e lemma it has a primitive $1$-form $k_u{\mathrm d} u+k_v{\mathrm d} v$, i.e. satisfying
\begin{equation}
\stackrel{2}{j}={\mathrm d}(k_u{\mathrm d} u+k_v{\mathrm d} v)=(\nabla_uk_v(u,v) -\nabla_vk_u(u,v)){\mathrm d}u\wedge{\mathrm d}v, 
\label{jfl}
\end{equation}
which proves~(\ref{jk}). The formulas for $k_{u,v}$ may be explicitly given in terms of $j$ as
\begin{equation}
[k_u(u,v),k_v(u,v)]=[-v,u]\,\int\limits_0^1 {\mathrm d}s\, s j(su,sv),\label{kukv}
\end{equation}
These quantities are not unequivocally defined, since they can be modified by adding $\partial_u \lambda(u,v)$ and $\partial_v \lambda(u,v)$ respectively for arbitrary scalar function $\lambda(u,v)$. This acts as a gauge operation for the current potential.

Let us now define the new magnetic field strength in the curved two-dimensional world:
\begin{equation}
\tilde{H}_u=H_u-k_u,\;\;\;\; \tilde{H}_v=H_v-k_v,
\label{Hnew}
\end{equation}
and, in accordance with~(\ref{BHuq}) and~(\ref{BHvq}), the induction
\begin{subequations}\label{Bnew}
\begin{align}
&\tilde{B}_u=B_u-\mu_0\sqrt{g_{3D}}\,(g_{3D})^{11}k_u,\label{Bnewu}\\
&\tilde{B}_v=B_v-\mu_0\sqrt{g_{3D}}\,(g_{3D})^{22}k_v,\label{Bnewv}
\end{align}
\end{subequations}
or equivalently
\begin{subequations}\label{BHf}
\begin{align}
&\stackrel{1}{\tilde{H}}=\stackrel{1}{H}-\stackrel{1}{k},\label{Hf}\\
&\stackrel{1}{\tilde{B}}=\stackrel{1}{B}-\mu_0*_2\stackrel{1}{k}.\label{Bf}
\end{align}
\end{subequations}

Upon this redefinition, the set of Maxwell equations~(\ref{mm}) takes the form
\begin{subequations}\label{tmm}
\begin{align}
&\nabla_u\tilde{H}_v-\nabla_v\tilde{H}_u=\partial_t D,\label{tmm1}\\
&\nabla_u E=\partial_t B_v+{j_{m}}_v,\label{tmm2}\\
&\nabla_v E=-\partial_t B_u-{j_{m}}_u,\label{tmm3}\\
&\nabla_u\tilde{B}_u+\nabla_v\tilde{B}_v=\rho_m,\label{tmm4}
\end{align}
\end{subequations}
where the components of the magnetic current are
\begin{subequations}\label{mc}
\begin{align}
\rho_m=&-\mu_0[\nabla_u(\sqrt{g_{3D}}\,(g_{3D})^{11}k_u)\nonumber\\
&+\nabla_v(\sqrt{g_{3D}}\,(g_{3D})^{22}k_v)],\label{mc1}\\
{j_{m}}_u=&\mu_0\sqrt{g_{3D}}\,(g_{3D})^{11}\partial_tk_u,\label{mc2}\\
{j_{m}}_v=&\mu_0\sqrt{g_{3D}}\,(g_{3D})^{22}\partial_tk_v,\label{mc3}
\end{align}
\end{subequations}
The magnetic charge density $\rho_m$ is not physical, and may be gauged away by imposing the appropriate conditions on $k_{u,v}$ (i.e. one can assume $\rho_m=0$). Therefore, we are left with a purely eddy magnetic current ${\mathbf j}_m$. 
The continuity equation is trivially satisfied, again due to the gauge condition:
\begin{equation}
\partial_t\rho_m+\nabla_u{j_{m}}_u+\nabla_v {j_{m}}_v=0.
\label{contm}
\end{equation} 

It would be interesting to visualize the flow of ${\mathbf j}_m$ in some special cases. This is done in Figure \ref{streams}. The stream lines on the plane and on the cylinder turn out to be quite natural. However, in the case of a truly curved surface, a cone, they behave differently. When curvature is very large, and this is the case close to the tip, the $\phi$ component of the current~(\ref{mc}) becomes enormous. This is a consequence of the fact that $(g_{3D})^{\phi\phi}$ tends to infinity. Because the stream lines cannot cross one another, all of them have to rotate around the cone axis.

The sources shown in Figure~\ref{streams} come from the scalar $j_m(u,v)$, which was chosen to be localized in space. For the case of surface-polarons sources they may be delocalized and are rather superpositions of those presented above. 

\section{Summary}\label{summary}

Summarizing the obtained results, one can say that the $2D$ theory of electromagnetism can be obtained from ordinary three-dimensional theory by the appropriate reduction to lower dimensional space in two possible ways. The first one  leads to ordinary electromagnetism with a two-component electric field and a scalar magnetic field with conserved electric currents flowing within the surface. It is the usual form to be found in textbooks. This theory is suitable for the approximate description of the $TE$ modes (i.e. those with electric field tangent to the surface) propagating within dielectric waveguides or along surfaces with evanescent tails in the outer space.

For the proper description of $TM$ modes (where magnetic field is tangent) one needs the other theory with the scalar $E$ and vector $\mathbf B$ fields. The appropriate reduction leading to this theory has been proposed. Then the reduced Maxwell equations contain scalar sources which may be subsequently rewritten in the form of eddy surface currents. This theory may have some application for example for the description of electromagnetic waves propagating in the interface between a conductor on the one side and a dielectric on the other. Due to the presence of the metallic surface only the $TM$ modes are admissible.

\section*{Acknowledgments}
The author would like to thank to Professors I. Bialynicki-Birula and J. Kijowski for interesting and elucidating discussions.
The work was supported by the Polish National Science Center Grant No. 2012/07/B/ST1/03347.

\end{document}